\documentclass{agujournal2019}
\usepackage{array}
\usepackage{graphicx}
\usepackage{tabularx}
\usepackage{booktabs}
\usepackage{amsmath}
\usepackage{amssymb}
\usepackage{multirow}
\usepackage{url} %this package should fix any errors with URLs in refs.
\usepackage{lineno}
\usepackage{soul}
\usepackage{bm}
\usepackage{overpic}
\usepackage{subfigure}

\newcommand{\Ek}{\mathrm{E}}
\newcommand{\Ra}{\mathrm{Ra}}
\newcommand{\Pran}{\mathrm{Pr}}
\newcommand{\Pm}{\mathrm{Pm}}
\newcommand{\Di}{\mathrm{Di}}
\newcommand{\Rey}{\mathrm{Re}}
\newcommand{\Rm}{\mathrm{Rm}}

\linenumbers
\draftfalse
\journalname{Journal of Geophysical Research: Planets}
\graphicspath{{images/}}
\nolinenumbers
\begin{document}
%% ------------------------------------------------------------------------ %%
%  Title
%
% (A title should be specific, informative, and brief. Use
% abbreviations only if they are defined in the abstract. Titles that
% start with general keywords then specific terms are optimized in
% searches)
%
%% ------------------------------------------------------------------------ %%

% Example: \title{This is a test title}

\title{Numerical Simulations of Magnetic Effects on Zonal Flows in Giant Planets}

%% ------------------------------------------------------------------------ %%
%
%  AUTHORS AND AFFILIATIONS
%
%% ------------------------------------------------------------------------ %%

% Authors are individuals who have significantly contributed to the
% research and preparation of the article. Group authors are allowed, if
% each author in the group is separately identified in an appendix.)

% List authors by first name or initial followed by last name and
% separated by commas. Use \affil{} to number affiliations, and
% \thanks{} for author notes.
% Additional author notes should be indicated with \thanks{} (for
% example, for current addresses).

% Example: \authors{A. B. Author\affil{1}\thanks{Current address, Antartica}, B. C. Author\affil{2,3}, and D. E.
% Author\affil{3,4}\thanks{Also funded by Monsanto.}}

\authors{Shanshan Xue\affil{1,2} and Yufeng Lin\affil{1}}

\affiliation{1}{Department of Earth and Space Sciences, Southern University of Science and Technology, Shenzhen 518055, PR China}
 \affiliation{2}{CAS Key Laboratory of Planetary Science, Shanghai Astronomical Observatory, Chinese Academy of Sciences, Shanghai, China}
% \affiliation{3}{Third Affiliation}
% \affiliation{4}{Fourth Affiliation}                                                                   

%(repeat as many times as is necessary)

%% Corresponding Author:
% Corresponding author mailing address and e-mail address:

% (include name and email addresses of the corresponding author.  More
% than one corresponding author is allowed in this LaTeX file and for
% publication; but only one corresponding author is allowed in our
% editorial system.)

% Example: \correspondingauthor{First and Last Name}{email@address.edu}

\correspondingauthor{Yufeng Lin}{linyf@sustech.edu.cn}

%% Keypoints, final entry on title page.

%  List up to three key points (at least one is required)
%  Key Points summarize the main points and conclusions of the article
%  Each must be 140 characters or fewer with no special characters or punctuation and must be complete sentences

% Example:
% \begin{keypoints}
% \item	List up to three key points (at least one is required)
% \item	Key Points summarize the main points and conclusions of the article
% \item	Each must be 140 characters or fewer with no special characters or punctuation and must be complete sentences
% \end{keypoints}

\begin{keypoints}
\item We show that the magnetic field suppresses zonal flows in the molecular layer and enhances convective motions.
\item We extract a scaling relation between the magnetic field and the zonal flow at the surface that matches observations of Jupiter and Saturn.
\item This study provides alternative evidence that the zonal winds on gas planets are deep-seated and convection driven.
\end{keypoints}

%% ------------------------------------------------------------------------ %%
%
%  ABSTRACT and PLAIN LANGUAGE SUMMARY
%
% A good Abstract will begin with a short description of the problem
% being addressed, briefly describe the new data or analyses, then briefly states the main conclusion(s) and how they are supported and
% uncertainties.

% The Plain Language Summary should be written for a broad audience,
% including journalists and the science-interested public, that will not have 
% a background in your field.
%
% A Plain Language Summary is required in GRL, JGR: Planets, JGR: Biogeosciences,
% JGR: Oceans, G-Cubed, Reviews of Geophysics, and JAMES.
% see http://sharingscience.agu.org/creating-plain-language-summary/)
%
%% ------------------------------------------------------------------------ %%
%% \begin{abstract} starts the second page
\justifying
\begin{abstract}
Jupiter and Saturn exhibit alternating east-west jet streams. The origin of these zonal flows has been debated for decades. The high-precision gravity measurements by the Juno mission and the grand finale of the Cassini mission have revealed that the observed zonal flows may extend several thousand kilometres deep and stop around the transition region from molecular to metallic hydrogen, suggesting the magnetic braking effect on zonal flows. In this study, we perform a set of magnetohydrodynamic simulations in a spherical shell with radially variable electrical conductivity to investigate the interaction between magnetic fields and zonal flows. A key feature of our numerical models is that we impose a background dipole magnetic field on the anelastic rotating convection. By varying the strength of the imposed magnetic field and the vigor of convection, we investigate how the magnetic field interacts with the convective motions and the convection-driven zonal flows. Our simulations reveal that the magnetic field tends to destroy zonal flows in the metallic hydrogen and suppress zonal flows in the molecular envelope, while the magnetic field may enhance the radial convective motions. We extract a quantitative relation between the magnetic field strength and the amplitude of zonal flows at the surface through our simulations, which roughly matches the observed magnetic field and zonal wind speed of Jupiter and Saturn. This discovery provides support from a new perspective for the scenario of deep convection-driven zonal winds which are confined to the molecular hydrogen layers in giant planets.

\end{abstract}
\setlength{\parindent}{0pt}
\section*{Plain Language Summary}
This study explores the mysterious east-west jet streams, known as zonal flows, observed on Jupiter and Saturn. Data from the Juno and Cassini missions suggested that these flows extend deep into the planets, stopping where hydrogen transitions from a molecular to a metallic state. Here we conduct computer simulations in a spherical shell geometry to understand how the magnetic fields and zonal flows interact in the deep interior of giant planets. We show that magnetic fields disrupt or weaken jet streams, but may enhance radial convective fluid motions. Crucially, our simulations provided a relation to link magnetic field strength with jet stream intensity at the surface, aligning well with real observations from Jupiter and Saturn.
This supports the theory that the jet streams on these giant planets are driven by deep convection and are confined to the outer molecular hydrogen layers. 
%% ------------------------------------------------------------------------ %%
%
%  TEXT
%
%% ------------------------------------------------------------------------ %%

\section{Introduction}
The distinct banded structures at the surfaces of Jupiter and Saturn are associated with alternating east-west winds. The broad eastward zonal jets around the equator can reach  a maximum velocity of about 150 m/s on Jupiter and about {350 m/s} on Saturn respectively {\cite{Tollefson2017,Garcia-Melendo2011}}. The strong equatorial jets are flanked by smaller-scale and weaker zonal jets with alternating directions. The formation mechanism and the penetration depth of these zonal winds have been studied for decades \cite{Ingersoll1969,Busse1983,Vasavada2005}. Broadly speaking, two scenarios have been proposed to explain the formation of the zonal winds in giant planets. The shallow scenario suggests that the zonal flows are formed in the top cloud layers as a weather phenomenon \cite{Ingersoll1969,Williams1978, Lian2008, Liu2010}, while the deep scenario suggests that the zonal flows are driven by deep-seated convection and extend into the molecular hydrogen layers \cite{Busse1983, Aurnou2001,Christensen2001}. It is also suggested that both of the aforementioned mechanisms are involved to account for observations, with the shallow-wind model applicable to the polar region and the deep-convection model to the equatorial region \cite{Yuan2023}.

The high-precision gravity measurements by the Juno mission and the grand finale of the Cassini mission indicate that the zonal flows extend to a depth of around 3000 km and 9000 km in Jupiter and Saturn respectively \cite{Kaspi2018,Galanti2019}.  {The penetration depths roughly coincident with, or are slightly shallower than, the region of the molecular-metallic phase transition} \cite{Wicht2019, Moore2019,Cao2023}.
However, the gravity inverse problem suffers from non-uniqueness and dynamical models used for the interpretation of the gravity signals remain debated \cite{Kong2018,Wicht2020}. If the zonal flows indeed extend to the semi-conducting region, magnetic field observations can provide additional constraints on the penetration depth of the zonal flows \cite{Duer2019,Galanti2021,wicht2024,Christensen2024}. It has been found that the variations of Jupiter's magnetic field are consistent with the advection of the zonal winds projected downward to the semi-conducting region, suggesting interaction between the magnetic field and zonal winds \cite{Moore2019}. The combined gravity–magnetic analysis reveals that the zonal flows are probably driven by deep convection and mainly confined to the weakly conducting and non-conducting outer layers \cite{Galanti2021}. However, the dynamical mechanism {of} the generation of zonal flows and interaction with the magnetic field {remains} to be elucidated.%This requires studies on magnetohydrodynamics (MHD) in the whole fluid layers within planets. 

Following the theoretical model by \citeA{Busse1983} on deep convection-driven zonal winds, several numerical models have been developed to simulate zonal jets driven by rotating convection in thin shells, modelling the hydrodynamics in the molecular envelope \cite<e.g.>{Aurnou2001, Christensen2001, Heimpel2005, Kaspi2009, Jones2009, Gastine2012}. These hydrodynamic simulations can produce strong prograde zonal flows around the equator with {a} few flanking zonal jets at high latitudes. Some recent hydrodynamic models can produce the observed features of both zonal flows and vortices \cite{Heimpel2016, Heimpel2022, Yadav2020}. However, hydrodynamic models assume an artificial boundary between the metallic and molecular hydrogen layers and typically impose non-penetration and stress-free conditions on the boundary, which may lead to missing some key physical processes for the deep convection model. The molecular-metallic transition does not lead to a sharp boundary and the electrical conductivity increases gradually with depth in giant planets \cite{French2012,Stevenson2020}. Therefore, the metallic region and the molecular region (including the cloud level) in giant planets form a continuously coupled fluid system. More importantly, the magnetic fields generated through dynamo action not only suppress zonal flows in the metallic region but also influence the overall convective dynamics in the whole fluid domain as we shall show in numerical simulations. It is thus necessary to study magnetohydrodynamics (MHD) in the whole fluid domain to study the formation mechanism of zonal flows if the {observed zonal jets at the surface are indeed driven by deep convection} \cite{Heimpel2011}.

The MHD in the interior of giant planets has been studied over the last decade motivated by understanding the magnetic field generation through dynamo simulations \cite<e.g.>{Jones2014, Glatzmaier2018, Duarte2018, Dietrich2018, Yuan2021}. These self-consistent dynamo simulations under the anelastic approximation have explored a wide range of control parameters using different reference states and electrical conductivity profiles. Some numerical models particularly focus on the interaction between the zonal flows and the dynamo action \cite{Heimpel2011,Gastine2014, Wicht2019,Yadav2022ApJ}. In dynamo simulations generating a strong dipole-dominated magnetic field, strong zonal flows are mainly generated outside the so-called magnetic tangent cylinder (MTC), {which takes the {electrical} conductivity transition radius as the cylinder radius, aligning with the axis of rotation.} It roughly separates the highly conducting and weakly conducting regions, {the common finding being that} zonal flows are suppressed inside the MTC. Recent studies suggested that a stably stratified layer in the semi-conducting region helps to produce multiple zonal jets and reconcile observations of the magnetic fields and zonal winds   \cite{Christensen2020,Gastine2021,Moore2022}. 

This study builds on previous dynamo simulations to further elaborate the magnetic effects on the zonal flows of giant planets. Different from the self-sustained dynamo models, we impose a dipole background magnetic field in anelastic convection-driven dynamos. 
{Imposing a dipole magnetic field allows us to control the magnetic strength in the model and thus explore the correlation between magnetic fields and zonal flows.}
%Note that the imposed dipole strength makes very little difference to the surface field, as can be seen by comparing $B_0$ and $B_{OB}$ in Table \ref{Tab:Models}. 
Meanwhile, we performed a self-sustained {dynamo} simulation without an imposed field for comparison. By varying the strength of magnetic field and the vigor of convection, we extract a scaling between the strength of the magnetic field and the amplitude of zonal flows at the surface. The derived scaling from numerical simulations roughly matches the observed magnetic field strength and the zonal wind speed at the surface of Jupiter and Saturn. Our numerical simulations provide further evidence for the scenario of deep convection driven zonal winds in giant planets.   
The remaining part of the paper is organized as follows. Section \ref{sec:num} introduces the numerical models and Section \ref{sec:Res} presents numerical results. The paper is closed by discussions and conclusions in Section \ref{sec:Con}.

\section{Numerical Models}\label{sec:num}
We consider a simplified model consisting of a rigid core of radius $r_i$ and a fluid shell of outer radius $r_o$ following the conventional view on the interior of giant planets. The radius ratio $\eta=r_i/r_o$ is fixed to be 0.3. The spherical shell uniformly rotates at $\mathbf {\Omega}=\Omega \mathbf{\hat z}$, and is filled with {fluid} of constant kinematic viscosity $\nu$ and thermal diffusivity $\kappa$, but with radially variable electrical conductivity $\sigma(r)$. Therefore, the magnetic diffusivity $\lambda=1/(\mu_0\sigma)$ also depends on the radius $r$, where $\mu_0$ is the vacuum magnetic permeability. We study the MHD (and hydrodynamics for comparison) in the rotating spherical shell driven by thermal convection under the anelastic approximation, which allows {a} background density variation with the radius but filters out fast-moving acoustic waves \cite{Braginsky1995,Lantz1999}. The MHD processes are described as perturbations of a hydrostatic reference state that is assumed to be adiabatic and spherically symmetric, so a thermodynamic variable can be written as $x(r,\theta,\varphi,t)=\overline{x}(r)+x'(r,\theta,\varphi,t)$, where the overbar denotes the reference state and the prime denotes a perturbation. Note that the dimensional reference state variables are denoted with {an} overbar, whereas the dimensionless reference state variables are denoted with {a} tilde superscript.

\subsection{Governing equations}
The dimensional momentum equation under the anelastic approximation is given as \cite<e.g.>{Jones2011} 
\begin{equation} \label{eq:NS_Dim}
\frac{\partial \mathbf u}{\partial t}+\mathbf{u \cdot \nabla u} +2\mathbf{\Omega \times u} = -\nabla \left( \frac{p'}{\overline{\rho}}\right)-\frac{s'}{c_p}\mathbf{g}
+\frac{1}{\overline{\rho}\mu_0}\mathbf{(\nabla \times B) \times B} 
 +\frac{\nu}{\overline{\rho}} \mathbf{\nabla \cdot} (2\overline{\rho} \mathbf{S}),
\end{equation}
with $\mathbf{u}$ the velocity, $\mathbf{B}$ the magnetic field, $\mathbf{g}$ the gravity, $\rho$ the density, $p$ the pressure, $s$ the specific entropy and $c_p$ the specific heat capacity at constant pressure. The strain-rate tensor $\mathbf S$ is defined as
\begin{equation}
\mathbf{S}=\frac{1}{2}\left[ \mathbf{\nabla u}+\left(\mathbf{\nabla u}\right)^{\mathrm T}\right]-\frac{1}{3}(\mathbf{\nabla \cdot u}) \mathbf{I},  
\end{equation}
with $^\mathrm{T}$ denotes the transpose and  $\mathbf{I}$ the identity matrix.

The mass conversation equation under the anelastic approximation is given as
\begin{equation} \label{eq:MassEq_Dim}
    \mathbf{\nabla \cdot} (\overline{\rho}\mathbf{u})=0. 
\end{equation}
The energy equation can be written as
\begin{equation} \label{eq:EnergyEq_Dim}
    \overline{\rho}\overline{T}\left( \frac{\partial s' }{\partial t}+\mathbf{u \cdot \nabla}s'\right)=\nabla \cdot \left( \kappa \overline{\rho}\overline{T} \nabla s' \right)+Q_\nu+Q_j,
\end{equation}
where $Q_\nu=2\overline{\rho} \nu \mathbf{S}^2$ and $Q_j=\lambda (\mathbf{\nabla \times B})^2/\mu_0$ represent the viscous and Ohmic heating respectively.

The magnetic induction {equation} with {variable} electrical conductivity reads
\begin{equation}\label{eq:MagInduction_Dim}
    \frac{\partial \mathbf{B}}{\partial t}=\mathbf{\nabla \times (u\times B)}+\mathbf{\nabla \times} (\lambda\mathbf{\nabla \times B}),
\end{equation}
and the magnetic field is divergence-free 
\begin{equation} \label{eq:DivB_Dim}
    \mathbf {\nabla\cdot B}=0.
\end{equation}

The above governing equations are solved in dimensionless form. Using the shell thickness $d=r_o-r_i$ as the length scale, the viscous {diffusion} time $d^2/\nu$ as the time scale, the entropy difference $\Delta s$ between the inner and outer boundaries as the entropy unit and the magnetic unit $\sqrt{\rho_{o} \mu_{0} \lambda_{i} \Omega}$ with $\rho_o$ the density at the outer boundary and $\lambda_i$ the magnetic diffusivity at the inner boundary, Equations (\ref{eq:NS_Dim}-\ref{eq:MagInduction_Dim}) in dimensionless form can be written as

\begin{equation}\label{eq:NS_Non}
        \frac{\partial \mathbf{u}}{\partial t}+\mathbf{u} \cdot \mathbf{\nabla}\mathbf{u} +\frac{2}{\Ek} \mathbf{\hat z} \times \mathbf{u}=-\nabla P+\frac{\Ra}{\Pran} \tilde{g} s' \mathbf{\hat r}+\frac{1}{\tilde{\rho}\Pm \Ek }(\mathbf\nabla \times \mathbf{B}) \times \mathbf{B}+\frac{1}{\tilde{\rho}} \mathbf{\nabla \cdot}(2\tilde{\rho}\mathbf{S}),
\end{equation}

\begin{equation}\label{eq:Mass_Non}
    \mathbf{\nabla} \cdot (\tilde{\rho} \mathbf{u})=0,
\end{equation}

\begin{equation}\label{eq:Energy_Non}
    \tilde{\rho} \tilde T\left(\frac{\partial s'}{\partial t}+\mathbf{u} \cdot \mathbf\nabla s'\right)=\frac{1}{\Pran} \mathbf\nabla \cdot(\tilde{\rho} \tilde{T} \mathbf\nabla s')+\frac{2\Pran \Di}{\Ra} \tilde{\rho} \mathbf{S}^2+\frac{\Pran\Di}{\Pm^2 \Ra \Ek}\tilde{\lambda}(\mathbf\nabla \times \mathbf{B})^2,
\end{equation}

\begin{equation}\label{eq:MagInd_Non}
    \frac{\partial \mathbf{B}}{\partial t}=\mathbf\nabla \times(\mathbf{u} \times \mathbf{B})-\frac{1}{\Pm} \mathbf\nabla \times(\tilde\lambda \mathbf\nabla \times \mathbf{B}).
\end{equation}

\begin{equation} \label{eq:DivB_Non}
    \mathbf {\nabla\cdot B}=0.
\end{equation}
Here $P$ is the reduced pressure including all potential terms and variables with {a} tilde represent normalized reference states. The density, temperature and radial gravity are normalized by the corresponding dimensional values ($\rho_o$, $T_o$, $g_o$) at the outer boundary, whereas the magnetic diffusivity is normalized by the dimensional value $\lambda_i$ at the inner boundary .
The non-dimensional control parameters are defined as
 \begin{equation}
     \Ek=\frac{\nu}{\Omega d^2},\quad \Ra=\frac{g_od^3\Delta s}{c_p\kappa\nu},\quad  \Pran= \frac{\nu}{\kappa}, \quad \Pm=\frac{\nu}{\lambda_i}, \quad \Di=\frac{g_o d}{c_p T_o},
 \end{equation}
 which represent the Ekman number, Rayleigh number, Prandtl number, magnetic Prandtl number and Dissipation number respectively. The dissipation number $\Di$ is not an independent control parameter. It is determined by the reference state as we shall show later. From now on, all physical quantities will be given in dimensionless form unless otherwise stated.

Equations (\ref{eq:NS_Non}-\ref{eq:DivB_Non}) are solved with the corresponding boundary conditions. For the velocity field, we use the stress-free condition at the outer boundary and the no-slip condition at the inner boundary. The entropy is fixed at both boundaries. For the magnetic field, we set {an} insulating boundary condition at both boundaries, meaning that the magnetic field matches a potential field. However, the axial dipole component at the inner boundary is fixed as the following
\begin{equation}\label{eq:B0}
    \mathbf{B}_{0}=B_0\left( \mathbf{\hat r} \cos \theta+\boldsymbol{\hat \theta}\frac{\sin \theta}{2} \right) \quad \mathrm{at} \quad r=r_i,
\end{equation}
where $B_0$ is the dimensionless magnetic field strength in Elsasser units. Different from self-sustained dynamo simulations, {this particular boundary condition is used to realize the background magnetic field.}

For the initial conditions, the velocity and entropy start from small random perturbations whereas the magnetic field starts from a dipole field which matches the inner boundary condition given in Equation (\ref{eq:B0}). {For the self-consistent dynamo case B2\_Self, the calculation started from a saturated state of the case B2\_20.}

\subsection{Reference States}
In order to simulate the MHD processes under the anelastic approximation, we need to prescribe a hydrostatic reference state of the fluid shell. In this study, we do not aim to model realistic interior of giant planets but rather focus on the MHD process based on simplified models. We assume the reference state is polytropic 
\begin{equation}
\tilde{p}=\tilde{\rho}^{1+\frac{1}{m}},
\end{equation}
 and in an ideal gas state 
 \begin{equation}
     \tilde p=\tilde{\rho} \tilde{T}.
 \end{equation}
Assuming the reference state is adiabatic and using the hydrostatic balance, one can obtain \cite{Jones2011}
\begin{equation}
    \frac{\mathrm d\tilde{T}}{\mathrm d r}=-\Di \tilde{g}.
    \label{eq:temp1}
\end{equation}
In order to uniquely determine the reference state from the above equations, we need to prescribe the radial gravity profile. {The gravity profile in Jupiter's interior increases with the radius in the deep region, and decreases with the radius in the outer region. In this study, we assume that the gravity is proportional to the radius, mostly characterising the deep interior, i.e. $\tilde{g}=r/r_o$ \cite{jones2000}.} The background temperature profile is then given by 
\begin{equation}
    \tilde{T}=\frac{\Di}{2r_o}\left(r_o^2-r^2\right)+1,
\end{equation}
and the density $\tilde \rho=\tilde T^m$. In the {above} simplified polytropic model, the dissipation number $\Di$ is determined by 
\begin{equation}
\Di=2\frac{\mathrm{e}^{N_\rho/m}-1}{1+\eta},
\end{equation}
where $m$ is the polytropic index $m$ and $N_\rho=\ln (\rho_i/\rho_o)$ is the number of density scale heights. 

 Ab-initio calculations reveal that the conductivity of Jupiter increases super-exponentially with depth in the molecular envelope, and smoothly varies in the metallic region \cite{French2012}. In this study, we use simplified electrical conductivity profiles following \citeA{gomez2010}
\begin{equation}
    \begin{split}
    \tilde{\sigma}=1+\left(\tilde{\sigma}_m-1\right)\left(\frac{r-r_i}{r_m-r_i}\right)^a \quad \text{for} \quad r<r_m, \\
    \\[0.06mm]
    \tilde{\sigma}=\tilde{\sigma}_m \exp\left[a\left(\frac{r-r_m}{r_m-r_i}\right)\frac{\tilde{\sigma}_m-1}{\tilde{\sigma}_m}\right] \quad \text{for} \quad r\geq r_m, 
    \end{split}
\end{equation}
where the conductivity $\tilde{\sigma}$ is normalized by the value at the inner boundary values, $\tilde{\sigma}_m$ corresponds to the conductivity at the transition radius $r_m$. 
The above formula provides a simplified representation of the realistic conductivity profiles of giant planets. The conductivity drops slowly with the radius when $r<r_m$ and then exponentially decays with the decay rate $a$ when $r>r_m$.

\begin{figure}[!ht]
    \centerline{\includegraphics[width=0.99 \textwidth]{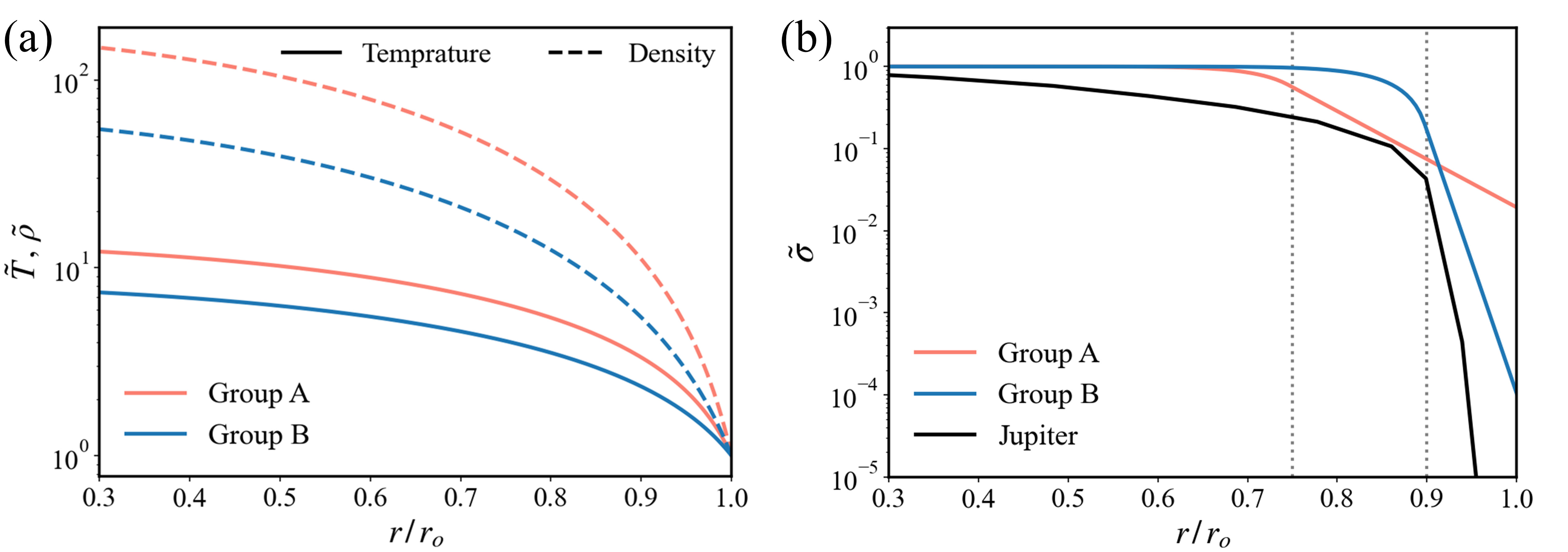}}
    \caption{Reference state profiles used in this study. (a) Density $\tilde{\rho}$ and temperature $\tilde T$ profiles. (b) Electrical conductivity $\tilde \sigma $ profiles. {The black profile indicates the conductivity variation in Jupiter's interior, which is normalised by the value at $0.09 R_J$ \cite{French2012}.} Vertical dashed lines correspond to the transition radius $r_m$.}
    \label{fig:ref}
\end{figure}

\begin{table}[t]
  \centering
  \setlength{\tabcolsep}{15pt}
  \caption{Physical parameters of reference states of the models.  }
  \begin{tabular}{cccccc}
    \toprule
    \specialrule{0em}{2pt}{1pt}
   Group   &  $m$  & $N_\rho$  & $\tilde{\sigma}_m$  & $r_m/r_o$  &  $a$ \\
   \specialrule{0em}{1pt}{2pt}
    \midrule
   Group A &  2  &  5 & 0.6 & 0.75 & 9 \\
   \specialrule{0em}{2pt}{2pt}
   Group B &  2  &  4 & 0.2 & 0.9 & 11 \\
    \bottomrule
  \end{tabular}
  \label{tab:ref}
\end{table}

In this study, we adopt two difference reference states and conductivity profiles. Numerical models in Group A have a thicker molecular envelope ($r_m/r_o=0.75$), while cases in Group B use a thinner molecular envelope ($r_m/r_o=0.9$).  The background profiles of density and temperature are shown in Figure \ref{fig:ref} (a) and the conductivity profiles are shown in Figure \ref{fig:ref}(b). {The density scale height chosen for Group B is slightly smaller than that of Group A in order to avoid the calculation errors caused by Alfv{\'e}n waves with excessively small scale.} Detailed parameters of the reference states and conductivity profiles are given in Table \ref{tab:ref}. We note that the outer boundary in our numerical models does not exactly correspond to the surface of giant planets because we cannot simulate huge density contrasts like giant planets. As we use the simplified polytropic model, it is difficult to define the equivalent depth in giant planets for the outer boundary in numerical models. {We assume that the outer boundary approximately corresponds to the radiative-convective boundary based on the background profiles.} 

\subsection{Numerical Method and Diagnostic Parameters}

Numerical simulations are implemented using the open-source pseudo-spectral code MagIC \cite{Gastine2012}, which is available at at \url{https://github.com/magic-sph/magic/}. The numerical method is based on an expansion of spherical harmonics in {the} angular direction and Chebyshev polynomials in the radial direction. The expansion is truncated to spherical harmonic degree $L$ and Chebyshev polynomial of degree $N$. Truncations used in our simulations are detailed in Table \ref{Tab:Models}. For the spherical harmonics transforms, the code makes use of the open-source library SHTns \cite{schaeffer2013}, which is available at \url{https://bitbucket.org/nschaeff/shtns}. For the time stepping, the explicit second-order Adams–Bashforth scheme is used for nonlinear terms and Coriolis forces, while the remaining terms are processed by {the} implicit Crank–Nicolson scheme \cite{Glatzmaier1984}. The numerical code has been benchmarked \cite{Jones2011} and widely used for simulating anelastic convection-driven MHD in spherical shells \cite{Wicht2019,Gastine2021}. 

Based on the dimensionless equations we used, the output velocity field and magnetic field are in units of the Reynolds number $\Rey$ and  and the square root of the Elsasser number $\Lambda$ respectively,
\begin{equation}
    \Rey=\frac{U d}{\nu}, \quad \Lambda=\frac{B^2}{\mu \lambda_i \rho_o \Omega},
\end{equation}
where $U$ and $B$ are dimensional velocity and magnetic field. In order to quantify the flow amplitude, we also made use of the Rossby number $\mathrm{Ro}$ as a global {diagnostic},
\begin{equation}
    \mathrm{Ro}=\frac{U_{rms}}{\Omega d},
\end{equation}
where $U_{rms}$ is the dimensional root-mean-squared velocity calculated from the kinetic energy. For the Rossby number, we define the zonal Rossby $\mathrm{Ro_{zon}}$ and the non-zonal Rossby $\mathrm{Ro_{non}}$, which correspond to the zonal flow (axisymmetric toroidal component) and non-zonal {flow} respectively.

\section{Numerical Results} \label{sec:Res}

\begin{table}[t] 
\renewcommand\arraystretch{1.3}
 \caption{Summary of the  control parameters and diagnostics of all simulations. For all cases we fix $\Pran=0.1$, $\Pm=1$ and $\eta=0.3$. For numerical truncations, we set $L=170$ and $N=161$. $OB$ indicates the quantities at the outer boundary.}
 \centering
 \setlength{\tabcolsep}{20pt}
 \tabcolsep=3pt
 \begin{tabular*}{\hsize}{@{}@{\extracolsep{\fill}}@{}cccccccccc}
 \midrule
    Case  &  $\Ek$  &  $\Ra$  &  $\Ra/\mathrm{Ra_c}$  &  $B_0$  &  $\mathrm{Ro_{zon}}$  &  $\mathrm{Ro_{zon}(OB)}$  &  $\mathrm{Ro_{non}}$  &  $\Lambda$ &  $B_{OB}$\\
 \hline
 \specialrule{0em}{3pt}{3pt}
    A1\_H & $1*10^{-5}$ & $3*10^8$ & 30 & -- & 0.1877 & 0.1343 & 0.0576 & -- & -- \\
    A1\_10  & $1*10^{-5}$ & $3*10^8$ & 30 & 10 & 0.0192 & 0.0939 & 0.0097 & 2.5529 & 0.0573  \\
    A1\_20  & $1*10^{-5}$ & $3*10^8$ & 30 & 20 & 0.0049 & 0.0253 & 0.0131 & 2.9492 & 0.8642  \\
     \specialrule{0em}{2pt}{2pt}
    A2\_H & $2*10^{-5}$ & $1.075*10^8$ & 25 & -- & 0.0304 & 0.1354 & 0.0099 & -- & --  \\
    A2\_20  & $2*10^{-5}$ & $1.075*10^8$ & 25 & 20 & 0.0244 & 0.1173 & 0.0104 & 0.6744 & 0.1471\\
     \specialrule{0em}{2pt}{2pt}
    A3\_H & $2*10^{-5}$ & $2.15*10^8$ & 50 & -- & 0.0438 & 0.1577 & 0.0200 & -- & -- \\  
    A3\_1  & $2*10^{-5}$ & $1.075*10^8$ & 50 & 1 & 0.0387 & 0.1410 & 0.0206 & 0.3312 & 0.0332 \\
    A3\_20  & $2*10^{-5}$ & $1.075*10^8$ & 50 & 20 & 0.0075 & 0.0299 & 0.0282 & 3.7471 & 1.0124 \\
    A3\_32  & $2*10^{-5}$ & $1.075*10^8$ & 50 & 32 & 0.0070 & 0.0283 & 0.0284 & 3.6062 & 1.0609 \\
     \specialrule{0em}{8pt}{8pt}
    B1\_H & $2*10^{-5}$ & $5.25*10^7$ & 15 & -- & 0.0208 & 0.0976  & 0.0057 & -- & -- \\ 
    B1\_5  & $2*10^{-5}$ & $5.25*10^7$ & 15 & 5 & 0.0134 & 0.0742 & 0.0065 & 3.0051 & 0.1933 \\
    B1\_20  & $2*10^{-5}$ & $5.25*10^7$ & 15 & 20 & 0.0111 & 0.0663 & 0.0079 & 2.2000 & 0.2081 \\
    B1\_50  & $2*10^{-5}$ & $5.25*10^7$ & 15 & 50 & 0.0104 & 0.0641 & 0.0079 & 3.2961 & 0.2596 \\
     \specialrule{0em}{2pt}{2pt}
    B2\_H & $2*10^{-5}$ & $1.05*10^8$ & 30 & -- & 0.0478 & 0.2013 & 0.0120 & -- & -- \\  
    {B2\_5} & {$2*10^{-5}$} & {$1.05*10^8$} & {30} & {5} & {0.0308} & {0.1331} & {0.0140} & {2.7637} & {0.1207} \\  
    {B2\_10} & {$2*10^{-5}$} & {$1.05*10^8$} & {30} & {10} & {0.0254} & {0.1123} & {0.0149} & {3.7900} & {0.1785} \\  
    B2\_20  & $2*10^{-5}$ & $1.05*10^8$ & 30 & 20 & 0.0110 & 0.0525 & 0.0198 & 7.7388 & 0.3843 \\
    {B2\_Self}  & {$2*10^{-5}$} & {$1.05*10^8$} & {30} & {-} & {0.0100} & {0.0496} & {0.0204} & {8.9788} & {0.7616} \\    
     \specialrule{0em}{2pt}{2pt}
    B3\_H & $2*10^{-5}$ & $1.75*10^8$ & 50 & -- & 0.0598 & 0.2136 & 0.0219 & -- & -- \\   
    B3\_20  & $2*10^{-5}$ & $1.75*10^8$ & 50 & 20 & 0.0113 & 0.0430 & 0.0317 & 13.5797 & 0.5981 \\
    \specialrule{0em}{3pt}{3pt}
 \midrule% 
 \end{tabular*}
 \label{Tab:Models}
 \end{table}

\begin{figure}[!ht]
    \centerline{\includegraphics[width=0.9 \textwidth]{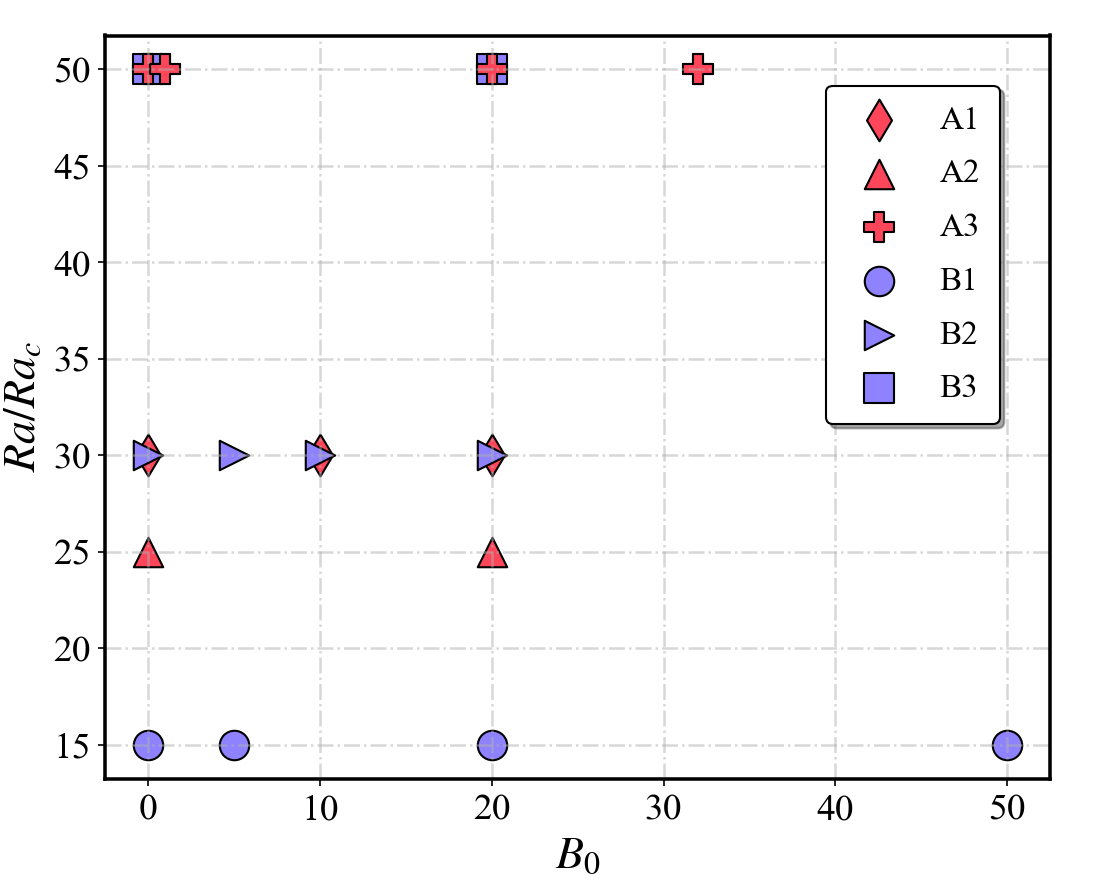}}
    \caption{Regime diagram of simulations in the parameter space of  the imposed field strength ($B_0$) and the supercriticality ($\Ra/\mathrm{Ra_c}$). Different colors and shapes represent simulations in different subgroups.}
    \label{fig:par}
\end{figure} 
In this study, we perform a set of numerical simulations to investigate the interaction between zonal flows and magnetic fields by varying the vigor of convection ($\Ra$) and the strength of the imposed magnetic field ($B_o$). Detailed control parameters of all simulations are listed in Table \ref{Tab:Models}. In all simulations, we fix the Prandtl number $\Pran=0.1$, the magnetic Prandtl number $\Pm=1.0$,  and the relative size of inner core $\eta=0.3$. Our simulations are divided into two groups according to the reference states and conductivity profiles used as we have mentioned. We further categorize simulations with the same Ekman number $\Ek$ and Rayleigh number $\Ra$ into subgroups A1, A2, A3 and B1, B2, B3. In each subgroup, we obtain the critical Rayleigh number $\mathrm{Ra_c}$ for hydrodynamic onset of convection. {Considering that the critical value $\mathrm{Ra_c}$ varies significantly with} different reference states and Ekman numbers, we use the ratio $\Ra/\mathrm{Ra_c}$ to measure the supercriticality of convection. For Group B2, we also show a self-sustained dynamo case $\mathrm{B2\_Self}$ to assess the impact of imposed fields. {However, the simulation of B2\_Self started from a saturated state of the B2\_20 case and evolved only for 0.01 viscous diffusion time due to computational limitations.} Figure \ref{fig:par} shows all simulations in the parameter space of the imposed field strength ($B_0$) and the supercriticality ($\Ra/\mathrm{Ra_c}$).  We find that our simulations in two groups show similar behaviors as we vary $B_0$ and $\Ra/\mathrm{Ra_c}$,  so our analysis in the following will focus on {numerical} models in the Group B. We will use all MHD simulations to extract the relationship between the magnetic field strength and the amplitude of zonal flow.

\begin{figure}[!ht]
    \centerline{\includegraphics[width=0.8 \textwidth]{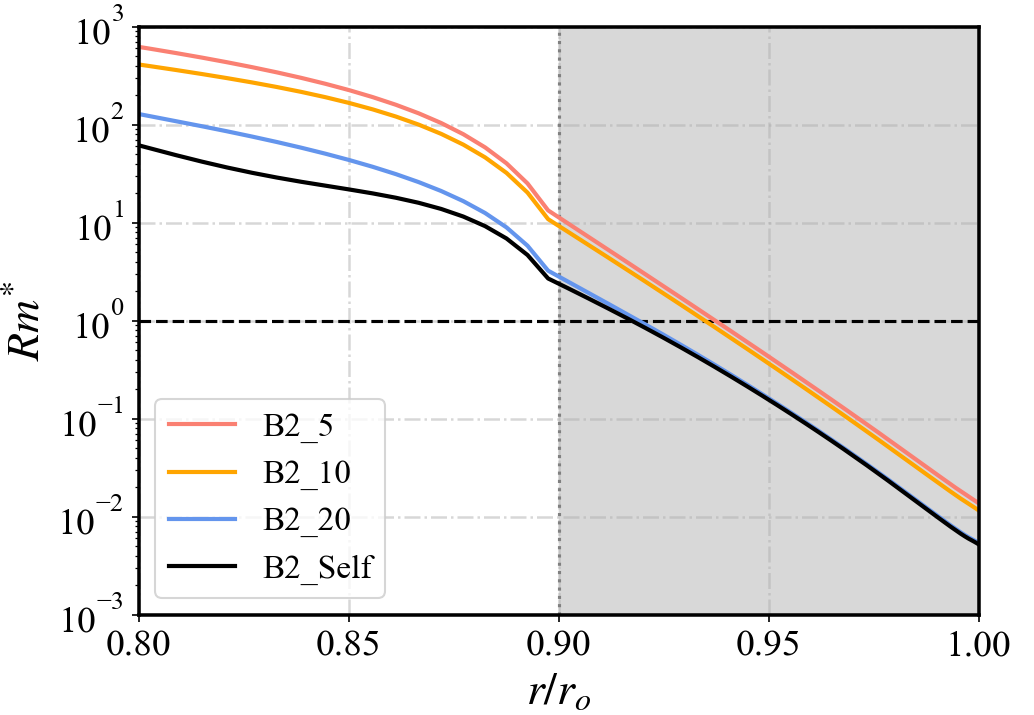}}
    \caption{The radial distribution of $\Rm^*$ outside $r=0.8$ for cases of Group B. The horizontal dashed line represents $\Rm^*=1$, and the gray area represents the molecular region. }
    \label{fig:rm}
\end{figure}

{To clarify the truncation depth in Jupiter's interior corresponding to the top border of our models, we introduce the modified magnetic Reynolds number $\Rm^*$ following \cite{Dietrich2018}, which is defined based on the the RMS zonal velocity and the magnetic diffusivity decay height $d_\lambda$, 
\begin{equation}
    \Rm^*(r) = \frac{U_{zon}(r) d_\lambda}{\lambda(r)}.
\end{equation}
Here $\lambda$ is the dimensional magnetic diffusivity and 
\begin{equation}
    d_\lambda=\left[\frac{1}{\lambda}\frac{ \mathrm{d}\lambda}{\mathrm{d}r}\right] ^{-1},
\end{equation}
is the magnetic diffusivity decay height. Although we have used the electrical {conductivity} profiles to define the metallic and hydrogen regions, the modified magnetic Reynolds number $\Rm^*$ provides a transition criterion between the electrically conducting and non-conducting region from the dynamic point of view \cite{Dietrich2018}. In the region where $\Rm^* <1$, the back-reaction of the magnetic field on the flow can be neglected and the magnetic field is approximately a potential field because there is no significant electrical current here. { Figure \ref{fig:rm} shows the modified Reynolds number $\Rm^*$ as a function of the radius for cases in Group B. Here we show the $\Rm^*$ only in the outer region because the modified Reynolds number is inappropriate in the inner region where the electrical conductivity remains constant in our models. 
We can see from figure \ref{fig:rm} that $\Rm^*$ decreases to the order of $O(10^{-2})$ at the outer boundary, suggesting that the outer region of the numerical models can be seen as the molecular envelope.}
However, the typical convection velocity deep inside Jupiter differs significantly from the zonal velocity at the surface by approximately an order of $O(10^3)$, whereas the maximum difference in our simulations is only in the order of $O(10)$. Reproducing such a significant velocity difference through numerical simulations is extremely challenging due to the computational limitations. However, our main focus is to investigate the correlation between the zonal flows in the molecular hydrogen layer and the internal magnetic fields, rather than attempting to reconstruct a Jupiter-like model through numerical simulations. For Group A, we found $\Rm^* \geq 1$, which implies that the outer boundary of numerical models in Group A remains in the weakly conducting region. Therefore, our following analysis will focus on cases in Group B.}

\subsection{Influence of the magnetic field}

\begin{figure}[!ht]
    \centerline{\includegraphics[width=1.0 \textwidth]{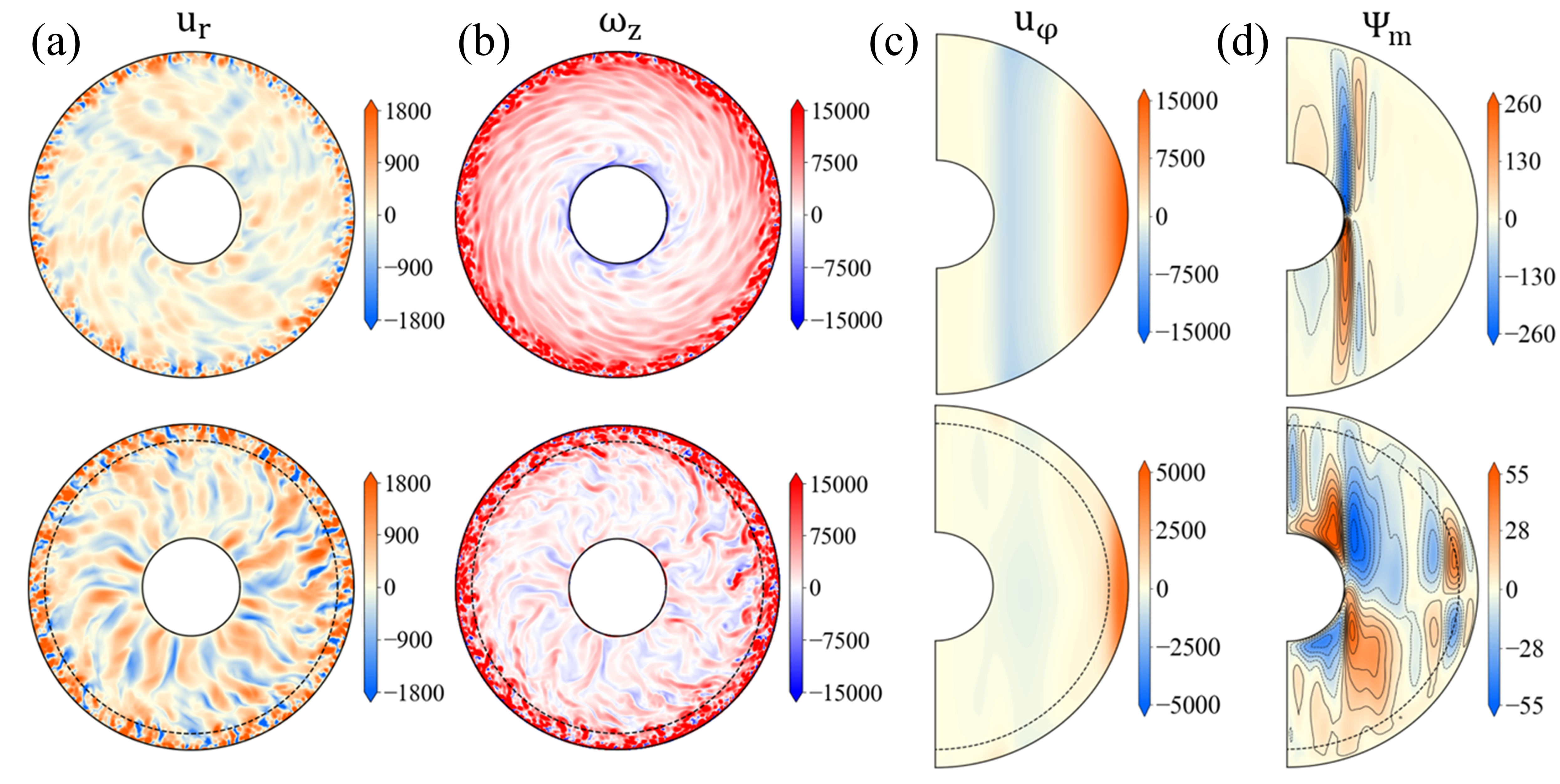}}
    \caption{Snapshots of the flow structure of two models, B2\_H (top panel) and B2\_20 (bottom panel).  (a) Radial velocity and (b) axial vorticity in the equatorial plane; (c) Time and azimuthally averaged $u_\varphi$ and (d) meridional circulation stream function $\Psi_m$ in the meridional plane. Dashed lines indicate the  marks the transition radius of the conductivity.}
    \label{fig:field}
\end{figure}

\begin{figure}[!ht]
    \centerline{\includegraphics[width=0.7 \textwidth]{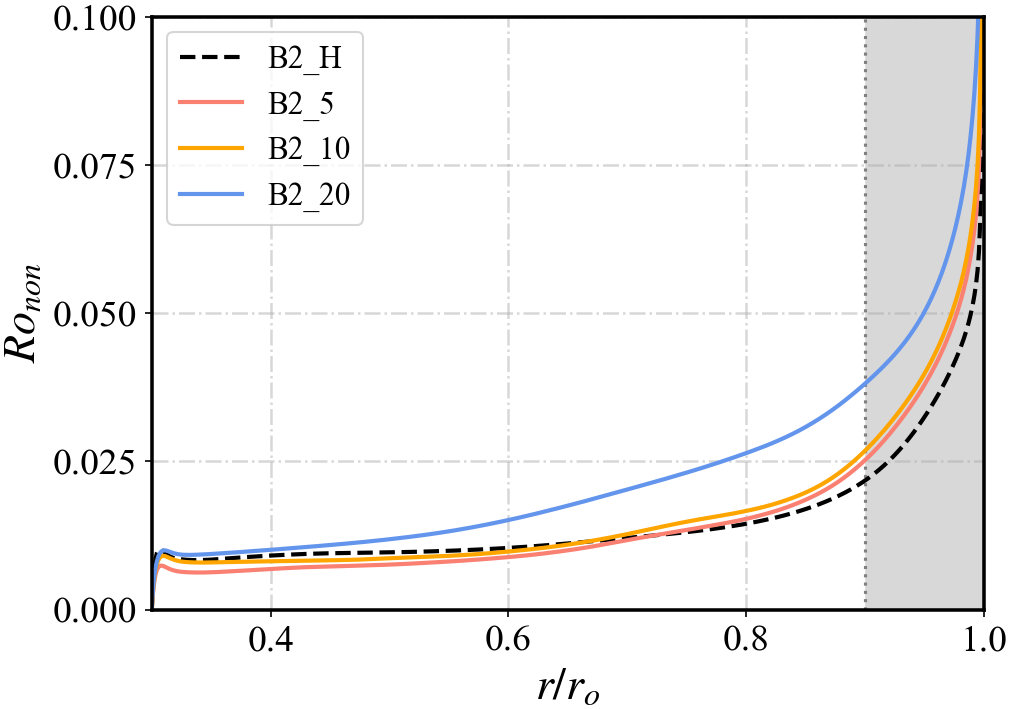}}
    \caption{{Non-Zonal Rossby number $\mathrm{Ro_{non}}$ versus radius for cases in B2. The gray area represents the molecular region.}}
    \label{fig:rnon}
\end{figure}

We use two cases in the subgroup B2 to illustrate the magnetic effect. The case B2\_H is a purely hydrodynamic model, whereas the case B2\_20 is a MHD model with a relatively strong imposed magnetic field of $B_0=20$. 
{Note that the dynamo cases in B2 exhibit similar flow patterns.} Figure \ref{fig:field} shows snapshots of the flow structure of the two models B2\_H (top panel) and B2\_20 (bottom panel).  Figure \ref{fig:field}(a) plots the radial velocity in the equatorial plane. In the hydrodynamic case, the convection tends to be more intensive in the thin outer envelope than in the deep interior.  In the MHD case, the convective motions are enhanced by the presence of a magnetic field in the metallic region. {This effect is more significant under the strong magnetic fields, as shown in Figure \ref{fig:rnon}. The Elsasser numbers of cases in B2 shown in Table \ref{Tab:Models} indicate that the total magnetic field intensity of the dynamo cases increases versus the imposed field strength. From Figure \ref{fig:rnon} we find that for case B2\_20, which exhibit the most significant magnetic fields, the spherically-averaged convective components are enhanced at all depths.} The enhanced convection by the magnetic field is also evident in the non-zonal Rossby numbers given in Table \ref{Tab:Models}. This likely resulted from the Lorentz force breaking some rotational constraints on the convection.

We note that flow structures and length scales are rather different between the metallic and molecular regions. The convective patterns are elongated in the radial direction by the magnetic field in the metallic region. The influence of the magnetic field on the convection flow pattern {can also be} seen from Figure \ref{fig:field}(b), which  shows the axial vorticity $\omega_z=(\boldsymbol\nabla \times \boldsymbol{u})_z$  in the equatorial plane. In the hydrodynamic case, the convection in the rotating shell is significantly affected by Coriolis force, showing mainly positive vorticity {in the outer region} and mainly negative vorticity near the inner boundary. This corresponds to strong zonal flows as shown in Figure \ref{fig:field}(c).  In the MHD case,  { the axial vorticity in the equatorial plane shows radially elongated structures with alternating positive and negative vorticity in the metallic region.} The {vorticity} does not change significantly with respect to the hydrodynamic model  in the molecular layer as expected.

Figure \ref{fig:field}(c) shows the mean zonal flow in the meridional plane, with red and blue representing the prograde and retrograde flows respectively. The hydrodynamic case shows classical quasi-geostrophic zonal flow field structure, namely a strong prograde jet in the equatorial region near the outer boundary, and a retrograde flow in the interior \cite{sreenivasan2006,miesch2006}. 
However, the zonal flow is significantly suppressed by the magnetic field in the MHD case both in the metallic and molecular regions. The prograde jet is basically confined outside the MTC. Figure \ref{fig:field}(d) shows the time and {azimuthally} averaged meridional circulation stream function $\Psi_m$, which is defined by $\tilde{\rho}\boldsymbol{u_m}=\boldsymbol{\nabla}\times (\tilde{\rho}\Psi_m\boldsymbol{e_\varphi})$, with  $\boldsymbol{u_m}$  the meridian velocity, and $\boldsymbol{e_\varphi}$  the unit vector of the azimuth direction. The solid and dashed lines  represent counterclockwise and clockwise circulations respectively. In the absence of the magnetic field, the meridional circulations are mainly adjacent to the tangent cylinder associated to the solid inner core. In the MHD case, meridional circulations are more divergent and span the entire shell.  In particular, we note that the meridional circulations can go across the transition between the metallic and molecular regions, leading to the exchange of momentum between two regions.
{There is also an important link between meridional circulation cells and alternating zonal flows in the middle and high latitudes{,} as found in shallow atmosphere observations analysis \cite{Duer2021}.} 

\begin{figure}
    \centerline{\includegraphics[width=0.7 \textwidth]{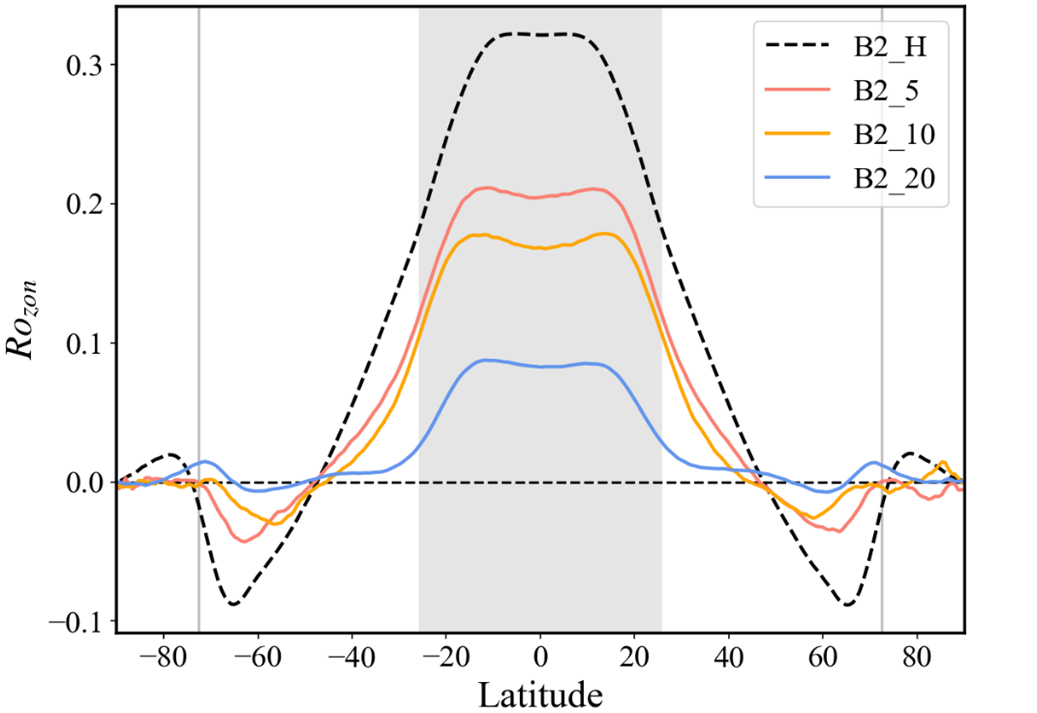}}
    \caption{Zonal Rossby number $\mathrm{Ro_{zon}}$ at the outer boundary as a function of the latitude for cases in B2. The gray area represents the molecular region, and the vertical grey line indicates the location of inner core tangent cylinder.}
    \label{fig:fieldzon}
\end{figure}

In order to see the zonal flow at the outer boundary more quantitatively, we plot the mean zonal Rossby number at the surface as a function of the latitude in Figure \ref{fig:fieldzon}. We can see that the magnetic fields have a significant effect on the zonal flows. The zonal flows velocities in the hydrodynamic simulation are higher than that in the MHD cases for all ranges of latitude, {and as the magnetic field strengthens, the zonal {flow} velocity become slower.} Besides, some weak alternating zonal flow bands occur at middle and high latitudes under the influence of magnetic fields. Compared with the non-magnetic case, the MHD simulation produces a slower prograde jet in the molecular layer. This suggests the magnetic field not only {brakes} the zonal flow in the metallic regions but also {reduces} the zonal flow in the molecular layer. Therefore, numerical models that treat the molecular layer as an isolated  thin shell lead to significant over-estimate of the generation of zonal flows. The overall flow morphologies are similar among simulations with different $\Ra$. 

\begin{figure}
    \centerline{\includegraphics[width=0.8 \textwidth]{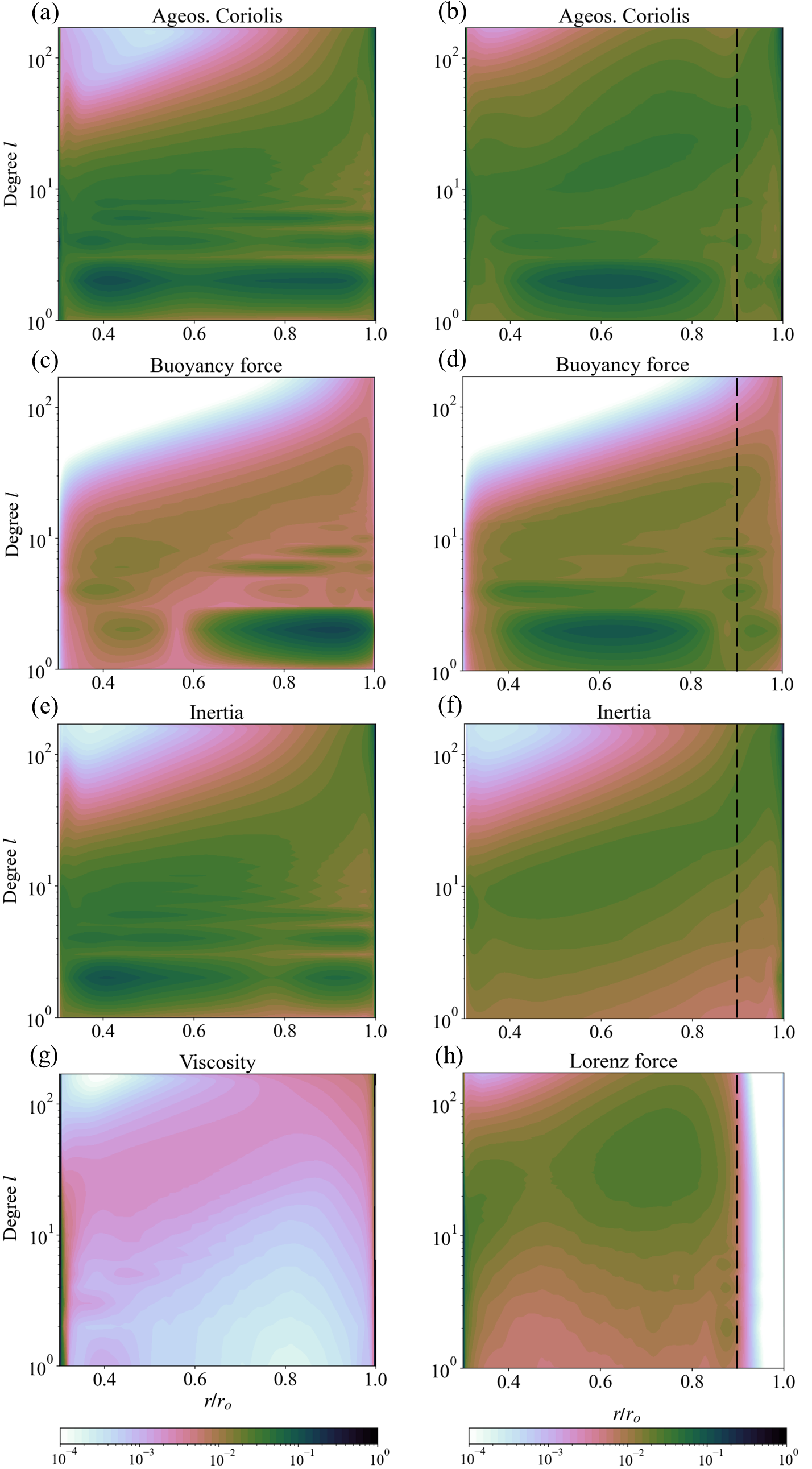}}
  \caption{Two-dimensional force balance spectra for the two cases, B2\_H (left panel) and B2\_20 (right panel). In each case, the spectra are normalized by the maximum force. The vertical dashed lines correspond to the electrical conductivity transition radius.  }
  \label{fig:fb}
\end{figure}

To further elaborate the magnetic effect on the dynamics,  Figure \ref{fig:fb} shows the 2-D force balance spectra in spherical harmonic degree $l$ and radius $r$ following \cite{Schwaiger2019}. The left panel shows the force balance of the hydrodynamic case B2\_ H and the right panel shows the MHD case B2\_20. We have removed the geostrophic Coriolis force balanced by the pressure gradient and show the non-geostrophic forces that are balanced by other force terms. In the hydrodynamic case, the buoyancy force is prominent in the outer region ($r/r_0>0.6$) at the large scale ($l \leq 3$). The {inertia} force is important at large and medium scales in most regions. The viscous force is small {compared} to other force terms. This shows that the hydrodynamic model is in the so-called CIA force balance after the leading order quasi-geostrophic balance as previous studies have shown.  In the MHD case, we show the Lorentz force instead of the viscous force. We can see that the presence {of a magnetic field enlarges} the contribution of the buoyancy force but makes the {inertia} force relatively weak. This is in line with Figure \ref{fig:field} showing that the magnetic field enhances the radial convective motions but suppresses the zonal flows. 
The Lorentz force clearly plays an important role in the metallic region in the MHD model.

\subsection{Dependence on the Rayleigh number}
\begin{figure}[!ht]
    \centering
    \begin{overpic}[width=14cm]{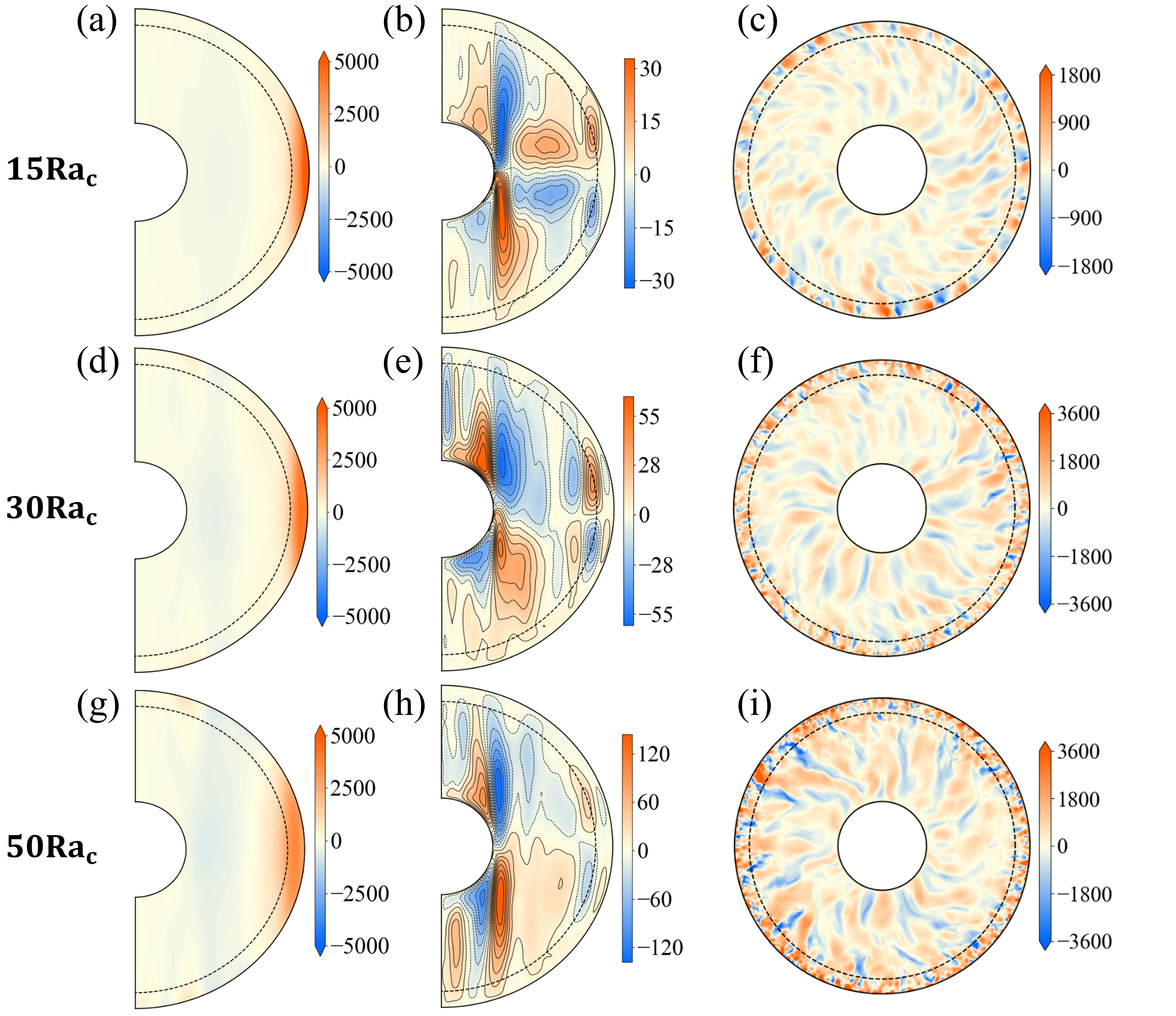}
    \end{overpic}
   \caption{Snapshots of the flow structure at different Rayleigh numbers. (a-c) B1\_20 model; (d-f) B2\_20 model; (g-i)  B3\_20 model. Left panel: mean zonal flow; Middle panel: mean meridional circulations; Right panel: radial velocity in the equatorial plane. The dashed lines indicate the  transition radius.}
    \label{fig:r4}
\end{figure}

We now analyze how the dynamics {change} as we {increase} the Rayleigh number $\Ra$. We focus on three models in Group B with the same imposed magnetic field $B_0=20$. Figure \ref{fig:r4} shows flow structures of three models at $\Ra=15\Ra_c$, $\Ra=30\Ra_c$ and $\Ra=50\Ra_c$. The overall morphologies of the flow structures are similar among three simulations at different $\Ra$. As we increase $\Ra$, convection becomes more vigorous as expected.   We also note from the middle panel that the meridional circulations become stronger {with} increasing $\Ra$. More vigorous convection and {larger} meridional circulations {with} increasing $\Ra$ are also reflected in the non-zonal Rossby numbers given in Table \ref{Tab:Models}. Although the imposed background magnetic field is the same, the diagnostic Elsasser number $\Lambda$ becomes larger {with} increasing $\Ra$, suggesting stronger magnetic fields are generated by more vigorous convection through dynamo actions.

\begin{figure}[!ht]
    \centerline{\includegraphics[width=1.0 \textwidth]{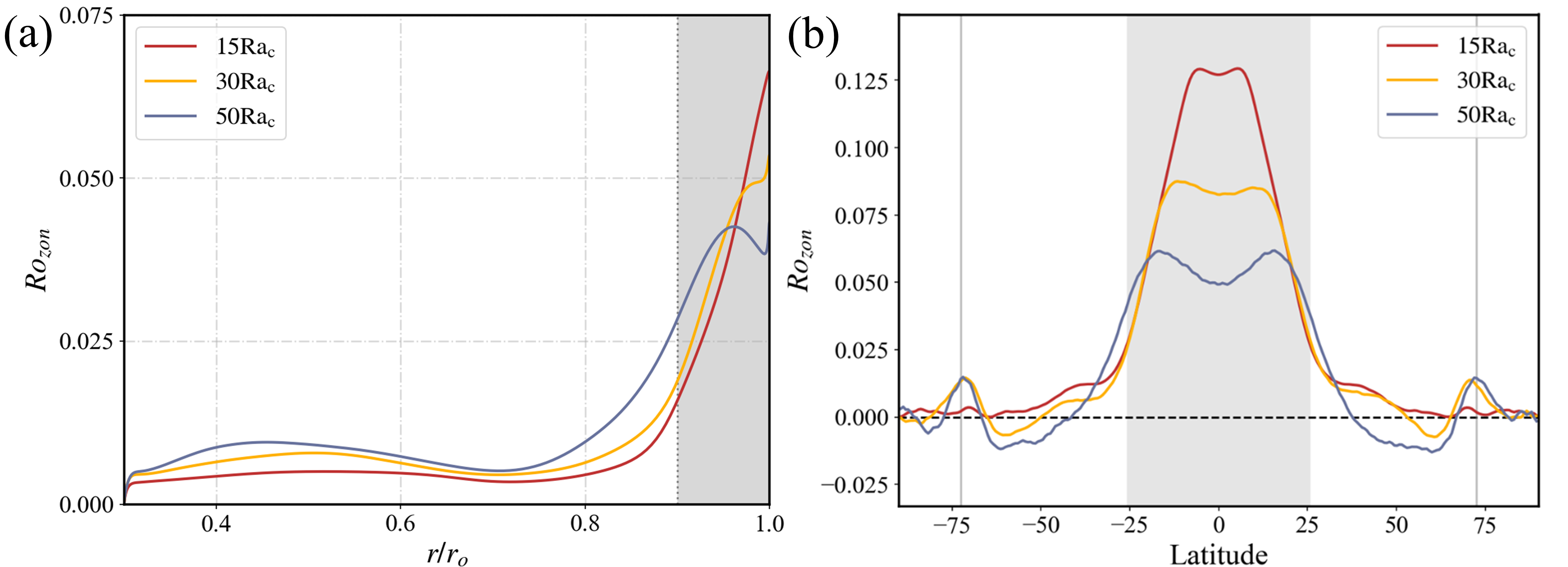}}
  \caption{Zonal Rossby number $\mathrm{Ro_{zon}}$ as a function of the radius (a) and the latitude at the surface (b) for three cases in Figure \ref{fig:r4}. Gray area represents the molecular region.}
  \label{fig:razon}
\end{figure}

The dependence of zonal flows on the Rayleigh number is more subtle as we can see from the left panel in Figure \ref{fig:r4} and from the zonal Rossby numbers in Table \ref{Tab:Models}. The global zonal Rossby number $\mathrm{Ro_{zon}}$ is nearly independent of $\Ra$, i.e. $\mathrm{Ro_{zon}}\approx 0.011$ in all three cases. However, the amplitude of zonal flows inside the MTC slightly increases {with} increasing $\Ra$ whereas the prograde zonal jet near the equator becomes weaker {as $\Ra$ is increased}. This feature can be seen more clearly in Figure \ref{fig:razon}(a) which plots $\mathrm{Ro_{zon}}$ as a function of the radius. Zonal flows inside and outside the MTC exhibit different behaviors {with} increasing $\Ra$ {suggesting} that the formation mechanisms of zonal flows are different. There are two possible mechanisms for driving zonal flows: one is caused by Reynolds stress resulting from statistical correlations between convective columns and the other is caused by thermal wind balance between buoyancy and Coriolis force \cite{aubert2005}. In the inner metallic hydrogen region, the quasi-geostrophic balance is destroyed by the magnetic field. Zonal flows are mainly generated by the thermal wind balance in strong dynamo regimes \cite{aubert2005}. When the Rayleigh number increases, the buoyancy force enhances the thermal wind effect and thus promotes the formation of zonal flow. In the outer molecular hydrogen layer, zonal flows are mainly driven by the Reynold stress which is probably weakened by strong buoyancy force at large $\Ra$, leading to weak zonal flows.

Figure \ref{fig:razon}(b) shows the zonal Rossby number at the outer boundary as a function of the latitude. The amplitude of zonal flows around the equator decreases significantly with the increase of Rayleigh number. However, increasing $\Ra$ tends to {promote} alternating zonal flows in the middle and high latitudes. We also notice that the zonal jets at low latitudes exhibit a valley at the equator, flanked by two maxima nearby in particular at large $\Ra$. {A similar} dimple is also seen in previous hydrodynamic simulations at sufficiently large $\Ra$ \cite{gastine2013}. It is interesting to note that {a} similar depression at the equator is observed in the Jupiter's zonal wind profile \cite{Porco2005}, but not on Saturn \cite{Porco2003}.

\subsection{Scaling relation between magnetic fields and zonal flows}
We have demonstrated the interaction between magnetic fields and zonal flows using a few cases. Generally speaking, the magnetic field tends to break the geostrophic constraint and suppress zonal flows. More importantly, the presence of magnetic field also {reduces} the zonal flow outside the magnetic tangent cylinder where the Lorentz force is vanishing. This is because the metallic and molecular regions are dynamically coupled. Therefore, we expect that observed zonal winds at the surface of giant planets should be related to the magnetic field which is generated in the metallic region through dynamo actions and can be observed above the surface. {In this section, we {analyse} the MHD simulations of Group B, which {incorporate} the metallic and the molecular hydrogen layer, to seek the relation between the amplitude of zonal winds and the intensity of magnetic fields at the surface.}
\begin{figure}[!ht]
    \centerline{\includegraphics[width=0.8 \textwidth]{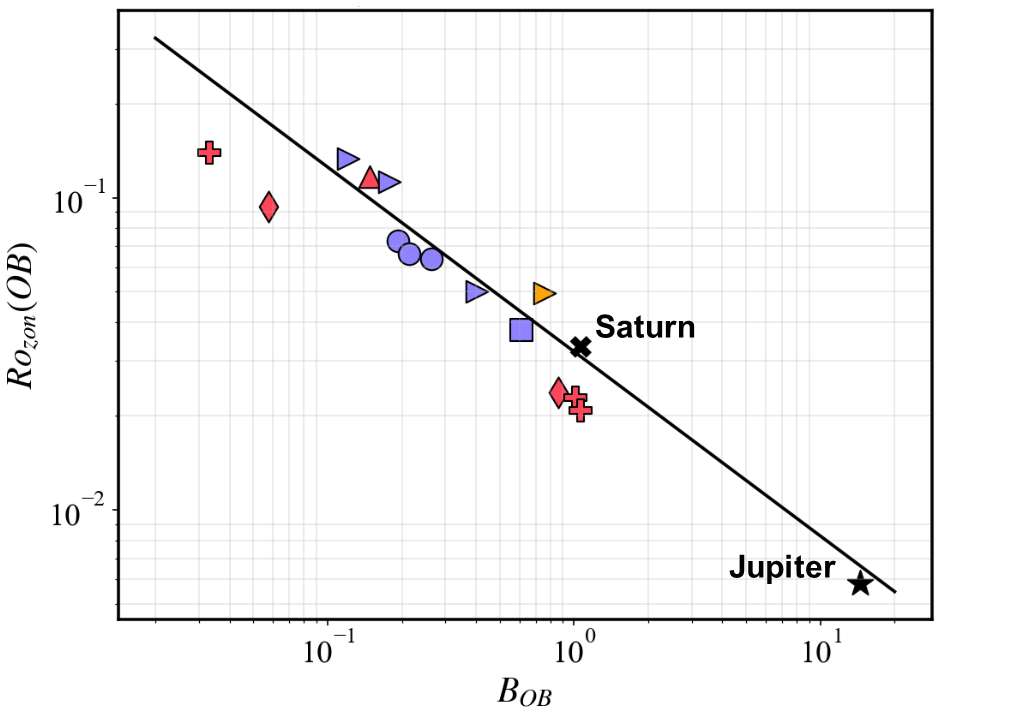}}
    \caption{The zonal Rossby number at the outer boundary $\mathrm{Ro_{zon}(OB)}$ as a function of the dimensionless magnetic field intensity (in Elsasser units) at the outer boundary $B_{OB}$ for MHD simulations. The black solid line corresponds to the least-square fitting of numerical data in Group B. Colors and shapes of symbols are the same as in Figure \ref{fig:par}. The orange triangle indicates the self-sustained dynamo case B2\_Self. Black symbols represent the estimated values of Jupiter and Saturn.}
    \label{fig:sca}
\end{figure}

\begin{table}[!ht]
  \centering
  \caption{Parameters used to estimate the $\mathrm{Ro_{zon}(OB)}$ and $B_{OB}$ for Jupiter and Saturn.
  $^1$\url{https://nssdc.gsfc.nasa.gov/planetary/factsheet/index.html} \\
  $^2$\citeA{nettelmann2012,guillot1999} \\
  $^3$\citeA{French2012,preising2023}\\
  $^4$\citeA{schubert2011}\\
  $^5$\url{https://atmos.nmsu.edu/planetary_datasets/indexwinds.html}.}
  \renewcommand{\arraystretch}{1.3}
  \setlength{\tabcolsep}{7mm}
  \begin{tabular}{ccc}
    \toprule
               &  Jupiter  &  Saturn \\
    \midrule
      $^1$$R$ (km) &  69911 & 58232 \\
      $^1$$T_\Omega$ (h) & 9.9259 & 10.656 \\
      $^2$$\rho_o$ ($\rm kg/m^3$) & 27.702 & 6.7563 \\
      $^3$$\sigma_i$ (S/m) & $3.39\times10^6$  & $1.60\times10^6$ \\
      $^4$Mean surface magnetic field ($\rm \mu T$) &  550  & 28 \\
      $^5$Mean zonal wind speed (m/s) & 49.748 & 222.27 \\ 
      $\mathrm{Ro_{zon}(OB)}$  &   0.006 &  0.033 \\ 
      $B_{OB}$ & 14.5 & 1.06 \\
    \bottomrule
  \end{tabular}
  \label{tab:js}
\end{table}

Figure \ref{fig:sca} plots the zonal Rossby number at the outer boundary $\mathrm{Ro_{zon}(OB)}$ versus the intensity of magnetic field at the surface $B_{OB}$ (in Elsasser units) for the MHD simulations and the estimated values of Jupiter and Saturn based on existing observations detailed in Table \ref{tab:js}. {Despite different control parameters being used in simulations, all numerical data points seem to follow a similar scaling relation between $\mathrm{Ro_{zon}(OB)}$ and $B_{OB}$ in {a} logarithmic scale. Therefore, we perform {a} least-square fitting of the numerical data points in Group B using a power law  
\begin{equation}
    \mathrm{Ro_{zon}(OB)}=K \left(B_{OB}\right)^\beta.
\end{equation}
 By doing so, we get the following scaling relation
\begin{equation}
    \mathrm{Ro_{zon}(OB)}=0.032{B_{OB}}^{-0.59},
    \label{eq:sca}
\end{equation}
which is shown as the black line in Figure \ref{fig:sca}. As we have discussed{,} the outer boundary of numerical models in Group A may not reach the non-conducting regime, {so} we use only numerical models in Group B to fit the scaling{.} We can see that the numerical data in Group A show {a} similar trend but may lead to a different scaling. This suggests that the detailed scaling depends on the reference states and electrical conductivity profile.
More systematic numerical simulations and theoretical justifications are required to refine the relation between the magnetic field {strength} and {the} zonal flow amplitude.} Nevertheless, it is very interesting that the observed zonal wind speeds and magnetic fields of Jupiter and Saturn at the surface roughly follow the scaling extracted from numerical simulations. Of course, the control parameters and reference states we used in numerical simulations are far from the realism of giant planets, and the shape of zonal wind profiles in simulations remains different from the observations. {Also, we use the observed surface wind speeds to estimate the Rossby number for Jupiter and Saturn, but it should be noted that the outer boundary of {our} numerical model {does} not exactly {correspond} to the surface of {the} giant planet.} Nevertheless, our numerical simulations suggest that the zonal wind {speed at the surface} is closely related to the strength of magnetic field generated in the deep interior. This provides numerical evidence that the zonal winds of giant planets are driven by deep-seated convection as revealed by gravity measurements of Juno and Cassini \cite{Kaspi2018,Galanti2019}.

\section{Conclusions} \label{sec:Con}
Motivated by understanding the generation mechanism of zonal flows in Jupiter and Saturn, we have performed a set of magnetohydrodynamic simulations in a rotating spherical shell with radially variable electrical conductivity model both {in} the metallic and molecular regions of giant planets. Fluid motions are driven by thermal convection under the anelastic approximation. Comparing to self-sustained dynamo simulations, a distinct ingredient of our numerical models is that we impose a background magnetic field by fixing the dipole component at the inner boundary, which allows us to adjust the strength of magnetic field more {flexibly}.   

By varying the strength of imposed magnetic fields and the vigor of convection, we investigate how the magnetic field interacts with the convective motions and the convection-driven zonal flows. Our simulations show that the magnetic field tends to destroy zonal flows in the conducting region as previous studies have shown. Meanwhile, the presence of a strong magnetic field may enhance the {radial convective motions} and meridional circulations by breaking the rotational constraint. In particular, the meridional circulations can cross the transition between the metallic and molecular regions, providing an exchange of momentum between {the} two regions. More importantly, we found that the magnetic field also {suppresses} zonal flows in the molecular envelope where the Lorentz force is vanishing. This suggests that the observed zonal winds at the surface would be closely related to the magnetic field if the zonal winds are deep-seated. We extract a scaling relation between the magnetic field strength and the amplitude of zonal flows at the surface from simulations as given in Equation \ref{eq:sca}. Strikingly, the observed magnetic field strength and zonal wind speeds at the surface of Jupiter and Saturn roughly follow the scaling relation extracted from numerical simulations. {This provides some new insights for understanding the origin puzzle of zonal flows in giant planets. Our numerical simulations provide support from a fresh perspective for the scenario of deep convection-driven zonal winds, which are roughly confined within the molecular hydrogen layers of giant planets.}

However, we should note that the control parameters and reference states used in numerical simulations are far from the {realistic values in} giant planets. The shape of zonal wind profiles in simulations is different from the observations of giant planets{, except for} the prograde jet around the equator. It remains a challenge for numerical models to produce multiple alternating zonal flows at high latitudes in a self-consistent manner as observed on Jupiter and Saturn. 
Nevertheless, our numerical simulations have demonstrated that the zonal wind speeds at {the} surface {are} closely related to the strength of {the} magnetic field that is generated in the deep interior. 

\section*{Data Availability Statement}
Numerical simulations made use of the open-source code MagIC \cite{Gastine2012}. The input parameters and diagnostic outputs of all simulations are given in Table \ref{Tab:Models}. 
%%%%%%%%%%%%%%%%%%%%%%%%%%%%%%%%%%%%%%%%%%%%%%%

\acknowledgments
The authors thank W. Dietrich and J. Wicht for helping to set the background magnetic field in the MagIC code, and D. Kong for helpful discussions and suggestions. We are very grateful to two anonymous reviewers for their constructive comments and suggestions. This study was supported by NSFC grants (42174215, 12250012), the B-type Strategic Priority Program of the CAS (XDB41000000) and the Pearl River Program (2019QN01X189). Numerical calculations were performed on the Taiyi cluster supported by the Center for Computational Science and Engineering of Southern University of Science and Technology.

%% ------------------------------------------------------------------------ %%
%% References and Citations

%%%%%%%%%%%%%%%%%%%%%%%%%%%%%%%%%%%%%%%%%%%%%%%
%
% \bibliography{<name of your .bib file>} don't specify the file extension
%
% don't specify bibliographystyle

% In the References section, cite the data/software described in the Availability Statement (this includes primary and processed data used for your research). For details on data/software citation as well as examples, see the Data & Software Citation section of the Data & Software for Authors guidance
% https://www.agu.org/Publish-with-AGU/Publish/Author-Resources/Data-and-Software-for-Authors#citation

%%%%%%%%%%%%%%%%%%%%%%%%%%%%%%%%%%%%%%%%%%%%%%%
\bibliography{xss_reference}

%Reference citation instructions and examples:
%
% Please use ONLY \cite and \citeA for reference citations.
% \cite for parenthetical references
% ...as shown in recent studies (Simpson et al., 2019)
% \citeA for in-text citations
% ...Simpson et al. (2019) have shown...
%
%
%...as shown by \citeA{jskilby}.
%...as shown by \citeA{lewin76}, \citeA{carson86}, \citeA{bartoldy02}, and \citeA{rinaldi03}.
%...has been shown \cite{jskilbye}.
%...has been shown \cite{lewin76,carson86,bartoldy02,rinaldi03}.
%... \cite <i.e.>[]{lewin76,carson86,bartoldy02,rinaldi03}.
%...has been shown by \cite <e.g.,>[and others]{lewin76}.
%
% apacite uses < > for prenotes and [ ] for postnotes
% DO NOT use other cite commands (e.g., \citet, \citep, \citeyear, \citealp, etc.).
% \nocite is okay to use to add references from your Supporting Information
%

\end{document}